\documentclass[aip,
               jcp,
               preprint,
               amsmath,
               superscriptaddress,
               floatfix,
               nobibnotes,
               ]{revtex4}

\usepackage{amssymb}
\usepackage{amsmath}
\usepackage{mathtools}
\usepackage{latexsym}
\usepackage{amsfonts}
\usepackage{bm}
\usepackage[dvips]{graphicx}
\usepackage{color}
\usepackage[utf8]{inputenc}
\usepackage[flushleft]{caption}
\usepackage{rotating}
\usepackage{txfonts}
\usepackage{booktabs}
\def\figfoot{CO/Cu, 21D MCCPD}
\DeclareMathAlphabet{\mathpzc}{OT1}{pzc}{m}{it}

\newcommand{\figcaption}[2]{
    \noindent {\bf Figure \ref{#1}:} #2
    \vspace{1cm}
}

\begin{document}


\title{High-dimensional Quantum Dynamics Study on Excitation-Specific Surface 
Scattering including Lattice Effects of a Five-Atoms Surface Cell}

\author{Qingyong Meng}
 \email{qingyong.meng@nwpu.edu.cn}
  \affiliation{Department of Chemistry,
               Northwestern Polytechnical University,
               West Youyi Road 127, 710072 Xi'an,
               China}

\author{Markus Schr{\"o}der}
 \email{markus.schroeder@pci.uni-heidelberg.de}
  \affiliation{Theoretische Chemie,
               Physikalisch-Chemisches Institut,
               Ruprecht-Karls Universit{\"a}t Heidelberg,
               Im Neuenheimer Feld 229, D-69120 Heidelberg,
               Germany}

\author{Hans-Dieter Meyer}
 \email{hans-dieter.meyer@pci.uni-heidelberg.de}
  \affiliation{Theoretische Chemie,
               Physikalisch-Chemisches Institut,
               Ruprecht-Karls Universit{\"a}t Heidelberg,
               Im Neuenheimer Feld 229, D-69120 Heidelberg, 
               Germany}

\date{\today}


\begin{abstract}

In this work high-dimensional (21D) quantum dynamics calculations on
mode-specific surface scattering of a carbon monoxide molecule on a 
copper (100) surface with lattice effects of a five-atom surface cell
are performed through the multilayer multiconfiguration time-dependent
Hartree (ML-MCTDH) method. We employ a surface
model in which five surface atoms near the impact site are treated
as fully flexible quantum particles while all other more distant atoms
are kept at fixed locations. To efficiently perform the 21D ML-MCTDH
wavepacket propagation, the potential energy surface is transferred to
canonical polyadic decomposition form with the aid of a
Monte Carlo based method. Excitation-specific sticking probabilities
of CO on Cu(100) are computed and lattice effects caused by the
flexible surface atoms are demonstrated by comparison with
sticking probabilities computed for a rigid surface.
The dependence of the sticking probability of the initial state
of the system is studied, and it is found that
the sticking probability is reduced, when the surface atom on
the impact site is initially vibrationally excited.
\end{abstract}

\maketitle

\section{Introduction\label{sec:intro}}

Surface scattering often leads to chemical bond ruptures, such that
in-depth studies of the gas-surface dynamics, in particular its
vibrational mode-specific features, are highly desirable. Molecular
vibration is well known to affect the reactive probability of the
gas-phase reactions, which can be predicted by the Polanyi rules
\cite{pol69:1439} for few-atom reactions \cite{xia11:440,yan15:60}.
In contrast to the gas phase, the gas-surface reactions are more 
complex, since molecular vibrational energy can be efficiently 
dissipated into the surface (or vice versa). Moreover, due to 
the fundamentally important role of the lattice motion in surface
scattering, the surface degrees of freedom (DOFs) must be included
when describing such processes in detail. This, of course, further
dramatically increases the complexity of the underlying models and
needs further development of the technology.
In the present contribution we study mode-excitation-specific scattering of a 
CO molecule on a Cu(100) surface with flexible five surface atoms
and compare it to scattering off a completely rigid surface model. To this 
end we employ a system-bath model where the CO molecule is treated as
the system while the surface DOF are regarded as the bath. 

Numerical calculations are performed using the Heidelberg implementation
of the multiconfiguration time-dependent Hartree (MCTDH) and its multilayer
extension (ML-MCTDH) 
\cite{mey90:73,man92:3199,bec00:1,mey09:book,mey12:351,wan03:1289,man08:164116,ven11:044135}.
To study mode-excitation-specific surface scattering, first ro-vibrational
eigen-states of the CO are computed at a fixed distance from the 
Cu surface using the block-relaxation algorithm of the MCTDH 
program \cite{mey90:73,man92:3199,bec00:1,mey09:book,mey12:351}. 
These eigen-states are then equipped with momentum towards the surface 
and propagated using the ML-MCTDH algorithm 
\cite{wan03:1289,man08:164116,ven11:044135}, where both, the system
and bath DOFs are treated in full dimensionality. Subsequently, sticking
probabilities and time-dependent expectation values of coordinates and
internal energies are computed. 

The efficiency
of the MCTDH and ML-MCTDH algorithms in the Heidelberg implementation
\cite{mey90:73,man92:3199,bec00:1,mey09:book,mey12:351,ven11:044135}
critically depends  from a representation of the Hamiltonian as a sum-of-products
(SOP) of operators that  exclusively depend on one coordinate (or a small
set of combined coordinates). In this work, we employ a so-called direct model 
for constructing the PES of the present molecule-surface problem. In
this model, all DOFs, surface and CO,  are included in the PES and treated 
on the same level. The direct model is very simple conceptionally,  but
the high dimensionality of the PES are often too large to build its SOP
form by potfit like methods\cite{jae96:7974,pel13:014108,sch17:064105}.
That is why we had developed an expansion model
\cite{men15:164310,men17:184305}, where part of the PES was
developed into a Taylor series.
Recently, Schr{\"o}der \cite{sch20:024108} proposed a Monte
Carlo method (denoted by MCCPD) to build a CPD form of the
high-dimension PES, making high-dimension dynamics calculations
by the direct model possible. Moreover, the the CP format (also called
CANDECOMP or PARAFAC in the literature) is more compact than the Tucker
format used in potfit related methods. Hence, the new MCCPD algorithm
does not only allow to re-fit higher dimensional potentials than before,
but produces SOP fits which much less terms than previously.
The CPU time consumed by a ML-MCTDH calculations scales almost
linearly with the number of SOP terms. 

The paper is organized as follows; in Section \ref{sec:theo}, we will
describe the theoretical framework, including the PES re-fitting
calculations and quantum dynamics calculations.
Section \ref{sec:numerical-details} gives the numerical
details of this work. Section \ref{sec:results} presents the flux
analysis results and time-dependent expectation values, 
together with discussions of the present results. Finally,
Section \ref{sec:con} concludes with a summary.

\section{Theoretical Framework\label{sec:theo}}

\subsection{Hamiltonian Model\label{sec:hamiltonian}}

\begin{figure}[h!]
  \centering
   \includegraphics[width=11cm]{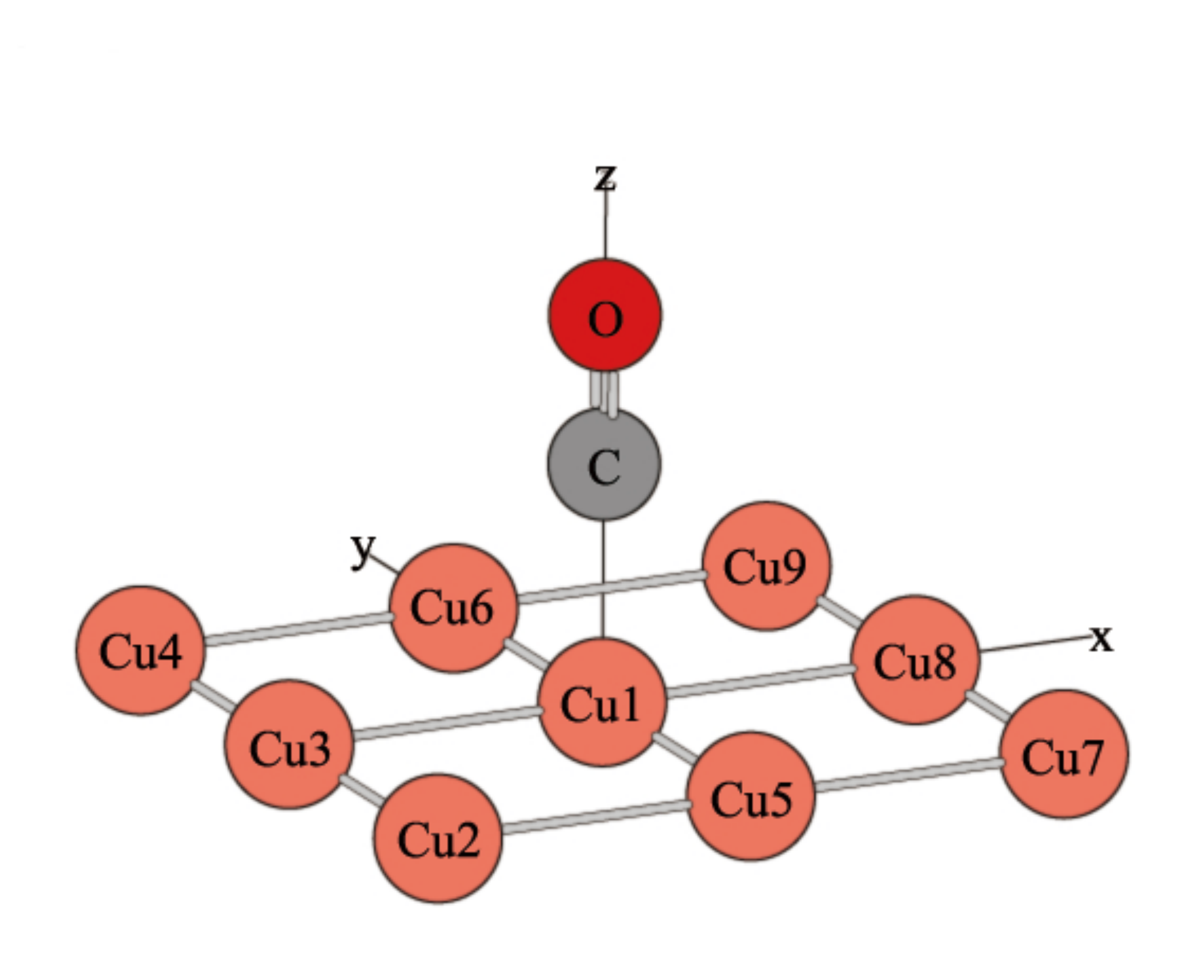}
    \includegraphics[width=11cm]{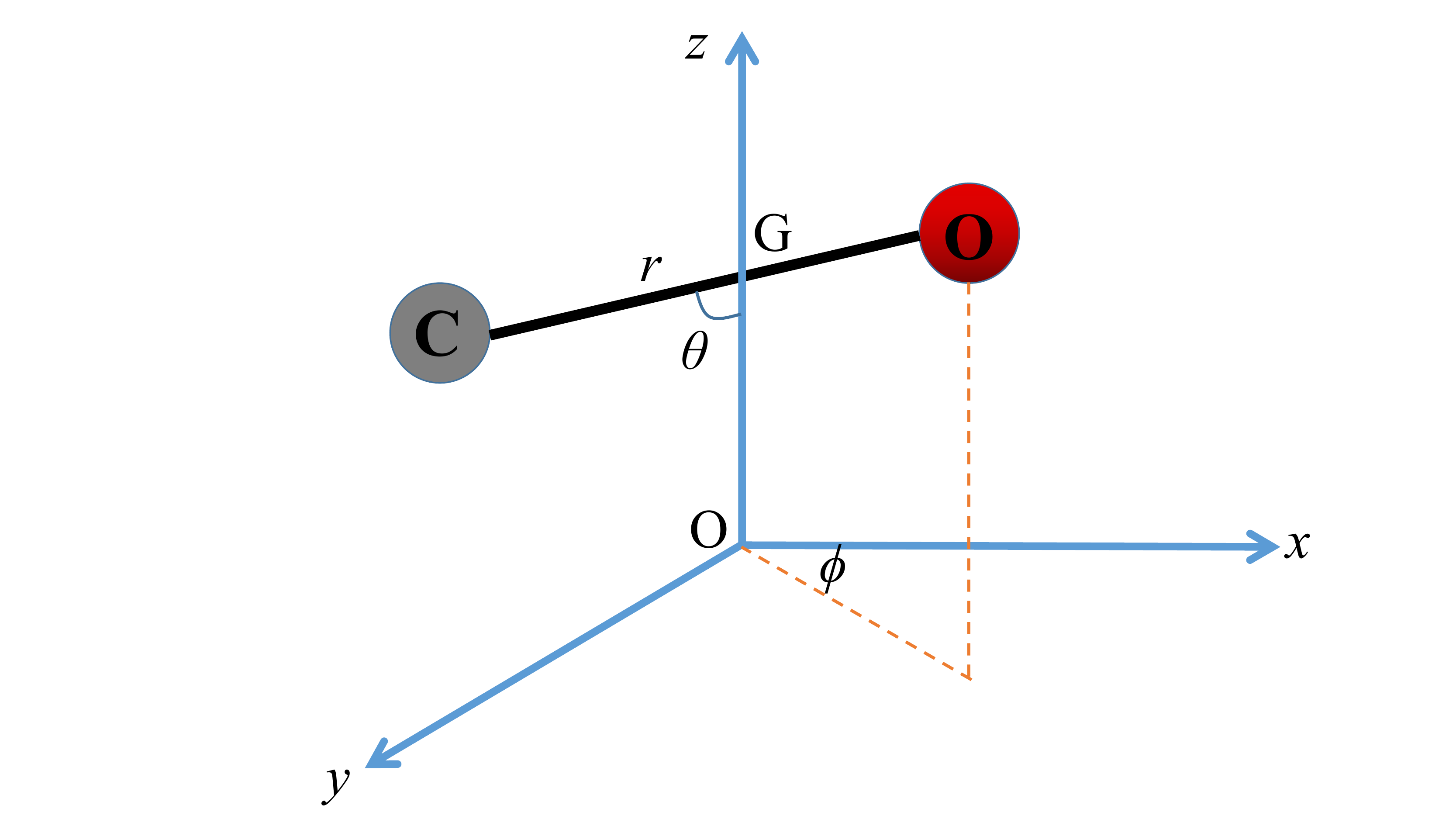}
     \caption{\figfoot} 
      \label{fig:geom}
       \end{figure}

Since the geometry of the CO/Cu(100) system and the expansion model
for the total Hamiltonian have been well discussed previously
\cite{men13:164709,men15:164310,men17:184305}, only a brief description
is given here. The coordinates used in this work are shown in Figure
\ref{fig:geom} (taken from reference \cite{men17:184305}). The center 
atom (called ``Cu1'') where the CO molecule will impact the surface as
well as its four neighboring surface atoms (labeled ``Cu3'', ``Cu5'',
``Cu6'', and ``Cu8'') are taken as dynamical particles, all other Cu 
atoms have fixed positions. As shown in Figure \ref{fig:geom}, the
coordinates of the five flexible Cu atoms are 
$\{\mathbf{Q}(1),\mathbf{Q}(3),\mathbf{Q}(5),\mathbf{Q}(6),%
\mathbf{Q}(8)\}=\mathbf{Q}$,
where $\mathbf{Q}(j)=\{X_j,Y_j,Z_j\}$ are Cartesian coordinates of the
$j$-th Cu atom. The coordinates describing the CO molecule 
above the surface are $\{x,y,z,r,\theta,\phi\}=\mathbf{q}$, where
$\{x,y,z\}$ denote Cartesian coordinates of the center-of-mass of the
CO molecule relative to the surface while $\{r,\theta,\phi\}$ denote
the C-O distance as well as rotation angles of CO relative to the
surface.

Having defined the coordinates in Figure \ref{fig:geom}, the total 
Hamiltonian $H(\mathbf{q},\mathbf{Q})$ of the present system-bath 
problem is partitioned as
\begin{equation}
H(\mathbf{q},\mathbf{Q})=T+V=T_s(\mathbf{q})+T_b(\mathbf{Q})+
V_b(\mathbf{Q})+V_{\mathrm{SAP}}(\mathbf{q},\mathbf{Q}), 
\label{eq:total-hamiltonian-operator}
\end{equation}
where $T_s$ and $T_b$ are the kinetic energy operators (KEO) of the CO 
molecule (system) and the copper surface model (bath), $V_b(\mathbf{Q})$ 
is the potential of the bath, while
$V_{\mathrm{SAP}}(\mathbf{q},\mathbf{Q})$ is the 21-dimensional SAP
PES \cite{mar10:074108,men13:164709,men15:164310,men17:184305} 
describing both, the internal system potential as well as the 
interaction between all system and all bath DOF. 

The 15-dimensional bath potential energy$V_b(\mathbf{Q})$ 
is modeled as a sum of Morse potentials between Cu atoms, 
$\mathcal{M}(\rho)=D[\exp(-2\alpha\rho)-2\exp(-\alpha\rho)]$. The Morse
interaction is in each Cartesian direction between nearest-neighbor 
atoms (including the fixed atoms, which by definition are fixed to
their equilibrium position). For the out-of-plane coupling between dynamical
surface atoms we used a harmonic interaction, because there is no 
directional preference. Then, the 15D bath potential $V_b(\mathbf{Q})$
can be written as \cite{men17:184305}
\begin{align}   
V_b(\mathbf{Q})=V_b(Q_{1x},\cdots,Q_{8z})
& =M(Q_{3x})+M(Q_{1x}-Q_{3x})+M(Q_{8x}-Q_{1x})+M(-Q_{8x})   \nonumber  \\
& +M(Q_{5y})+M(Q_{1y}-Q_{5y})+M(Q_{6y}-Q_{1y})+M(-Q_{6y})    \nonumber  \\
& +M(Q_{5x})+M(-Q_{5x})+M(Q_{6x})+M(-Q_{6x})                  \nonumber  \\
& +M(Q_{3y})+M(-Q_{3y})+M(Q_{8y})+M(-Q_{8y})                   \nonumber  \\
& +M(Q_{1z})+M(Q_{3z})+M(Q_{5z})+M(Q_{6z})+M(Q_{8z})            \nonumber  \\
& +\frac{k}{2}(Q_{1z}-Q_{3z})^2+\frac{k}{2}(Q_{1z}-Q_{5z})^2
+\frac{k}{2}(Q_{1z}-Q_{6z})^2+\frac{k}{2}(Q_{1z}-Q_{8z})^2.
\label{eq:pot-bath}
 \end{align}
The parameters in $V_b(\mathbf{Q})$, {\it i.e.}, $D$, $\alpha$, and 
$k$ were obtained
\cite{men17:184305} by fitting the vibrationally excited energies of
a clean Cu(100) surface to theoretical \cite{cor01:235118,hei03:151,bor08:075428} 
and experimental \cite{wut86:71,wut86:445,che91:11394,hei03:151} 
vibrational energies such that the spectral properties of the bath 
model system resemble those of a realistic surface. Note that the 
bath potential is already in SOP form as required for the numerical 
calculations later.

According to Figure \ref{fig:geom}, the KEO of the CO system is written as (atomic units, $\hbar=1$)
\begin{equation} 
T_s=-\frac{1}{2m}\left(\frac{\partial^2}{\partial x^2}
+\frac{\partial^2}{\partial y^2}
+\frac{\partial^2}{\partial z^2}\right)
-\frac{1}{2\mu}\frac{\partial^2}{\partial r^2}+\frac{1}{2mr^2}J^2(\theta,\phi),  
\label{eq:keo-sys}
\end{equation}
where $m$ is the mass of the CO molecule and $\mu$ its reduced mass. 
The symbol $J^2(\theta,\phi)$ denotes the operator for the squared 
angular momentum operating on $\theta$ and $\phi$, that is 
\begin{equation}
J^2(\theta,\phi)= -\frac{1}{\sin\theta}\frac{\partial}
{\partial\theta}\left(\sin\theta \frac{\partial}
{\partial\theta}\right)-\frac{1}{\sin^2(\theta)}
\frac{\partial^2}{\partial\phi^2}.
\label{eq:angle-p}
\end{equation}
Moreover, the KEO of the bath is defined as 
\begin{equation} 
T_b=-\frac{1}{2M}\sum_j\left(\frac{\partial^2}{\partial X_j^2} 
+\frac{\partial^2}{\partial Y_j^2} 
+\frac{\partial^2}{\partial Z_j^2}\right),
\label{eq:keo-bath}
\end{equation}
where $M$ is the mass of one Cu atom. The summation in Equation \eqref{eq:keo-bath}
covers all flexible copper atoms in Figure \ref{fig:geom}. 

\subsection{The MCCPD Algorithm\label{sec:mccpd-algorithm}}

The efficiency of the MCTDH algorithm in the Heidelberg implementation
derives from a representation of the Hamiltonian as a SOP form of
the Hamiltonian operator 
\begin{equation} 
H = \sum_r\prod_\kappa^fh_r^{(\kappa)}(q_\kappa)
\label{eq:hamilton}
\end{equation}
where the constituent operators $h_r^{(\kappa)}(q_\kappa)$ exclusively
depend on one (physical or logical, {\it i.e}. combined) coordinate $q_\kappa$.
Since the KEO given in Equations \eqref{eq:keo-sys} and \eqref{eq:keo-bath} 
as well as the bath potential Equation \eqref{eq:pot-bath} are already in 
SOP form, only the 21D potential function,
$V_{\mathrm{SAP}}(\mathbf{q},\mathbf{Q})$, must be recast into
a SOP form. To this end we use a recently proposed method
\cite{sch20:024108} for fitting a high-dimensional PES into a canonical
polyadic decomposition (CPD) form, also known as PARAFAC or
CANDECOMP in the literature. 

The CPD form of the $f$-dimensional potential function $V(q_1,\cdots,q_f)$
can be written as \cite{sch20:024108}
\begin{equation} 
V(q_1,\cdots,q_f)\simeq V^{\mathrm{CPD}}(q_1,\cdots,q_f)=
\sum_{r=1}^Rc_rv_r^{(1)}(q_1)\cdots v_r^{(f)}(q_f),
\label{eq:cpd-potential-form-00}
\end{equation}
where $R$ is the expansion order, also called the rank of the CPD
expansion, while $\{q_1,\cdots,q_f\}$ are the $f$  (physical or logical)
coordinates. The expansion basis functions $v_r^{(\kappa)}(q_{\kappa})$ 
in Equation \eqref{eq:cpd-potential-form-00}, the so-called single-particle 
potentials (SPP), exclusively depend on one coordinate. It is assumed that
the SPP are normalized, {\it i.e.},
\begin{equation}
\left\langle v_r^{(\kappa)}(q_{\kappa})\big\vert 
v_r^{(\kappa)}(q_{\kappa})\right\rangle=1,
\label{eq:cpd-potential-form-01}
\end{equation}
but otherwise no further restrictions are imposed. It is in particular
not required that the basis functions are orthogonal. Numerically, the
SPP are sampled on an underlying time-independent primitive basis, 
typically grid points, such that Equation \eqref{eq:cpd-potential-form-00} 
can be re-cast into a grid-based form. Here the one-dimensional SPP
function $v_r^{(\kappa)}(q_{\kappa})$ is replaced by its values evaluated
on grid points, where $q_{\kappa,i_\kappa}$ is the $i_\kappa$th sampling
point of the coordinate $q_{\kappa}$, and 
$v_{r,i_{\kappa}}^{(\kappa)}=v_{r}^{(\kappa)}(q_{\kappa,i_\kappa})$
is the SPP evaluated at this point. Then, the multidimensional potential
function can be written in a tensors notation as
\begin{equation}
V_I^{\mathrm{CPD}}=\sum_rc_r\prod_{\kappa}v_{r,i_{\kappa}}^{(\kappa)}
=\sum_rc_r\Omega_{r,I},
\label{eq:cpd-potential-form-02}
\end{equation}
where the multi-index $I=(i_1,\cdots,i_f)$ and the definition
$\Omega_{r,I}=\prod_{\kappa}v_{r,i_{\kappa}}^{(\kappa)}$
have been used. 

The remaining task is to find both the expansion functions and
the coefficients. To this end, one starts with a set of functionals 
\begin{equation}
\mathcal{J}_\kappa=\sum_I W_I^{\kappa}\left(V_I-V_I^{\mathrm{CPD}}\right)^2
+\epsilon\sum_rc_r^2\sum_I W_I^{\kappa}\Omega^2_{r,I},
\label{eq:cpd-potential-form-03}
\end{equation}
for each mode $\kappa$, where 
$W^{\kappa}_I=1_\kappa\sum_{i_\kappa}W_I = 1_\kappa W_{I^\kappa}^{\kappa}$ 
is a positive and coordinate-dependent
weight function $W_I$ where the $\kappa$th DOF has been integrated out
and replaced by unity, and $\epsilon$ is a regularization parameter,
typically set to square root of machine precision. 
Here, in passing, we introduced the index
$I^{\kappa}=(i_1,\cdots,i_{\kappa-1},i_{\kappa+1},\cdots,i_f)$
which is the full combined index with the $\kappa$th sub-index missing. 
Note that the first part of Equation \eqref{eq:cpd-potential-form-03}
measures the difference of the CPD fit to the exact potential subject
to the weight function, while the second part is called 
the regularization for reasons that will become obvious later. It is
introduced to penalize for (almost) linearly dependent terms in the
CPD expansion, which may arise due to ill-conditioned matrices in the
minimizing algorithm. 
The weight function will serve two purposes later: on the one hand
it will be used to emphasize regions of interest where increased fitting
accuracy is required. These will be the low energy regions where the
wavefunction resides. On the other hand, the weight function will serve
as a distribution function of sampling points when later the complete
sum over $I$ is replaced by Monte-Carlo sampling. 

To find the minimum of Equation \eqref{eq:cpd-potential-form-03} one performs
the the functional derivative of $\mathcal{J}_\kappa$ with respect to one SPP
and coefficient of the coordinate $q_\kappa$ and obtains \cite{sch20:024108}
\begin{equation}
\frac{\delta\mathcal{J}_\kappa}{\delta c_rv_{r,i_\kappa}^{(\kappa)}}=
-2\sum_{I^{\kappa}}W^{\kappa}_{I_\kappa}V_I\Omega_{r,I^{\kappa}}^{\kappa}
+2\sum_{r'}c_{r'} v_{r',i_\kappa}^{(\kappa)}S_{r,r'}^{\kappa}
+2\epsilon c_r v_{r,i_\kappa}^{(\kappa)}S_{r,r}^{\kappa}=0,
\label{eq:cpd-potential-form-04}
\end{equation}
where the abbreviations
\begin{equation} 
S_{r,r'}^{(\kappa)}=\sum_{I^{\kappa}} W^{(\kappa)}_{I_\kappa}
\Omega_{r,I^{\kappa}}^{\kappa}\Omega_{r',I^{\kappa}}^{\kappa},
\label{eq:cpd-potential-form-05}
\end{equation}
and
\begin{equation} 
\Omega_{r,I^{\kappa}}^{\kappa}=\prod_{\kappa'\neq\kappa}
v_{r,i_{\kappa'}}^{(\kappa')}
\label{eq:cpd-potential-form-06}
\end{equation}
have been used. From Equation \eqref{eq:cpd-potential-form-04}, 
a linear equation can be found in the form \cite{sch20:024108}
\begin{equation} 
\sum_{I^{\kappa}}W^{\kappa}_{I_\kappa}V_I\Omega_{r,I^{\kappa}}^{\kappa}
=\sum_{r'}S_{r,r'}^{(\kappa)}\left[1+\epsilon\delta_{r,r'}\right]
c_rv_{r,i_{\kappa}}^{(\kappa)}, 
\label{eq:cpd-potential-form-07}
\end{equation}
that can be solved with standard linear algebra tools. This is the working 
equations of the present PES re-fitting process.
As the solutions of the working equations \eqref{eq:cpd-potential-form-07} 
depend on the solutions of all other DOF one can now iteratively optimize
by solving Equation \eqref{eq:cpd-potential-form-07} for each mode and 
using the obtained solution in the following optimization. This scheme 
is called alternating least squares (ALS). 
One should notice, however, that other than in the traditional ALS, the form given
above will usually not lead to a monotonically increasing fit accuracy unless
the weight function is separable \cite{sch20:024108}. In practice, however,
the fit accuracy will usually increase monotonically for an initial number of
iterations before it starts to (mildly) fluctuate at which point one stops
the optimization.

The main bottleneck in solving the working equations \eqref{eq:cpd-potential-form-07}
is that it contains multi-dimensional quadratures over $I^{\kappa}$ that in general 
cannot be separated into products of lower-dimensional integrals. If the 
dimensionality of the system becomes too large, these integrals cannot
be completely evaluated any more. To overcome this, the complete sums 
are replaced by a Monte-Carlo integration, where the weight function serves as 
the distribution function of the sampling points $\left\{s^\kappa\right\}$ 
which are a sample drawn from the set of all quadrature points 
$\left\{I^\kappa\right\}$. With this, Equation \eqref{eq:cpd-potential-form-07}
becomes
\begin{equation}
\sum_{s^{\kappa}} V_{i_\kappa s^\kappa}\Omega_{r,s^{\kappa}}^{\kappa}
=\sum_{r'}Z_{r,r'}^{(\kappa)}\left[1+\epsilon\delta_{r,r'}
\right]c_rv_{r,i_{\kappa}}^{(\kappa)}, 
\label{eq:cpd-potential-form-08}
\end{equation}
with
\begin{equation}
Z_{r,r'}^{(\kappa)}=\sum_{s^{\kappa}} 
\Omega_{r,s^{\kappa}}^{\kappa}\Omega_{r',s^{\kappa}}^{\kappa}.
\label{eq:cpd-potential-form-09}
\end{equation}
Note, that in Equation \eqref{eq:cpd-potential-form-08} the primitive
grid of the mode $\kappa$ is not subject to Monte-Carlo sampling such that 
this index is complete. We refer the reader to reference 
\cite{sch20:024108} for further technical details on MCCPD.

\subsection{The MCTDH and ML-MCTDH Algorithms\label{sec:mctdh}}

In the present contribution we use the
Heidelberg implementation \cite{mctdh:MLpackage} of the MCTDH
algorithm, more precisely the multi-layer variant (ML-MCTDH)
\cite{wan03:1289,man08:164116,ven11:044135,wan15:7951} for solving
the time-dependent Schr{\"o}dinger equation. ML-MCTDH is particularly 
suited for treating high-dimensional systems as in the present case.
The algorithm is well discussed in the literature, such that we only 
give a brief introduction here.

\begin{figure}[h!]
  \centering
   \includegraphics[width=24cm,angle=90]{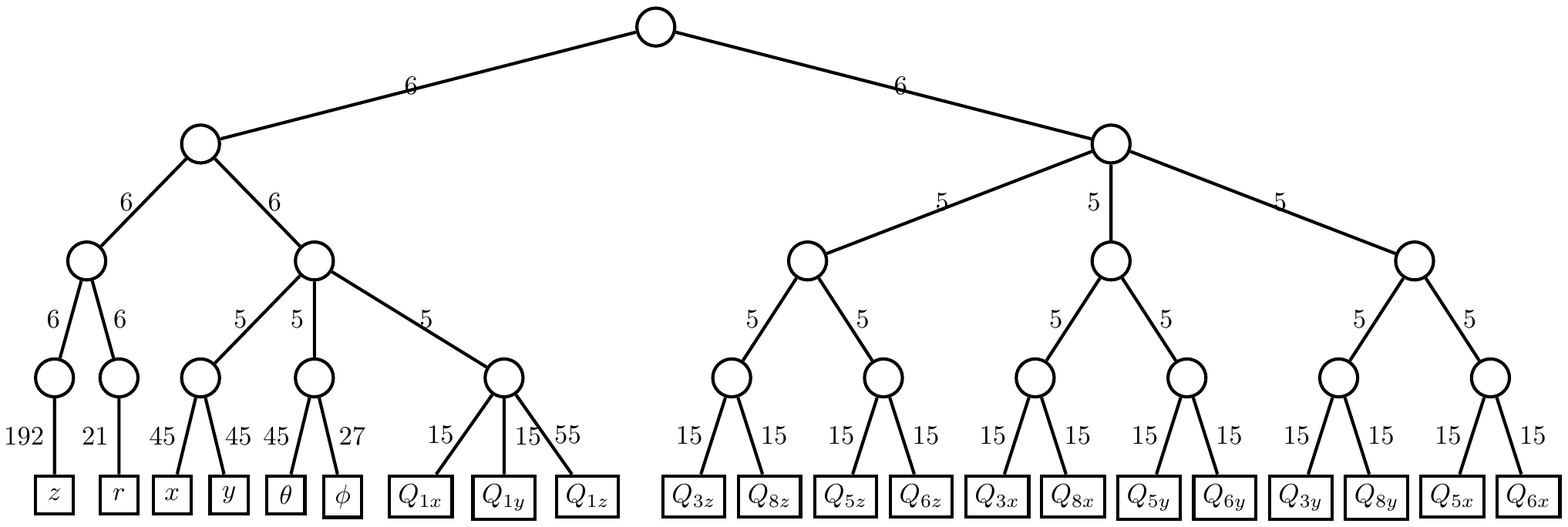}
    \caption{\figfoot}
     \label{fig:MLtree}
      \end{figure}

Within the ML-MCTDH algorithm the total 
time-dependent nuclear wavefunction is expressed in terms of a tensor
in a hierarchical Tucker format which has a tree-like structure.
To this end the wavefunction is expanded in a set of 
multi-dimensional, time-dependent basis functions, also called 
single particle functions (SPFs), which are themselves expanded in an 
underlying multi-dimensional, time-dependent basis as outlined in 
equation \eqref{eq:ml-wf}. This scheme is repeated until in the
lowest level a time-independent primitive basis is used.
The expansion can be written as
\begin{equation}
\varphi_{m}^{z-1;\kappa_{l-1}}(Q_{\kappa_{l-1}}^{z-1},t) =
\sum_{j_{1}}^{n_1^z}\cdots\sum_{j_{p_{\kappa_l}}}^{n_{\kappa_l}^z}
A_{m;j_{1},\cdots,j_{p_{\kappa_l}}}^{z}(t)\prod_{\kappa_l=1}^{p^z}
\varphi_{j_{\kappa_l}}^{z,\kappa_l}(Q_{\kappa_l}^z)
= \sum_JA_{m;J}^z\cdot\Phi_J^z(Q_{\kappa_{l-1}}^{z-1}),
\label{eq:ml-wf}
\end{equation}
where $z=\{l;\kappa_1,\cdots,\kappa_{l-1}\}$ and
$z-1=\{l-1;\kappa_1,\cdots,\kappa_{l-2}\}$. The symbol $l$ denotes
the layer depth and $z$ indicates a particular node in the ML-tree.
The complete nuclear wave function $\Psi$, which is to be
identified with $\varphi^0$, is expanded by the time-dependent SPF
with $l=1$. The logical coordinate,
$Q_{\kappa_{l-1}}^{z-1}=\big\{Q_1^z,\cdots,Q_{p_{\kappa_l}}^z\big\}$,
is a combination scheme of underlying coordinates $Q_i$. At the
bottom layer the SPFs are to be replaced with time-independent primitive
basis functions. The structure of an multi-layer wavefunction is most
conveniently visualized by a plot of the aforementioned tree structure.
The tree used in the present work is shown in Figure \ref{fig:MLtree}.

By inserting the multi-layer Ansatz, equation \eqref{eq:ml-wf}, into the 
Dirac-Frenkel variational principle, the ML-MCTDH equations of motion
(EOM) for arbitrary layering schemes have been derived together with 
an algorithm for the recursive evaluation of all intermediate quantities
entering the ML-MCTDH EOM \cite{wan03:1289,man08:164116,ven11:044135}.
According to Wang and Thoss \cite{wan03:1289}, Manthe \cite{man08:164116},
and Vendrell and Meyer \cite{ven11:044135}, the ML-MCTDH EOM have a
very similar structure to the usual MCTDH equations, and for the top
layer coefficients they are identical to the MCTDH ones, that is
\begin{equation}
i\frac{\partial A_{I}^{1}}{\partial t}
=\sum_{J}\Big\langle\Phi_{I}^{1}\Big\vert\hat{H}
\vert\Phi_{J}^{1}\Big\rangle A_{J}^{1}
\end{equation}
where the top layer configurations
\begin{equation}
\Phi_{J}^{1}=\varphi_{j_{1}}^{1;1}(Q_{1}^{1},t)
\cdots\varphi_{j_{p}}^{1;p}(Q_{p}^{1},t),
\end{equation}
are defined as direct products of SPFs and the multi-index 
$J=j_{1},\cdots,j_{p}$ has been implicitly introduced. The
EOM for the propagation of the SPFs are formally the same 
for all layers
\begin{equation}
i\frac{\partial\varphi_{n}^{z,\kappa_{l}}}{\partial t}
=\Big(1-P^{z,{\kappa_l}}\Big)\sum_{j,m}
\Big(\rho^{z,\kappa_{l}}\Big)_{nj}^{-1}\cdot
\Big\langle\hat{H}\Big\rangle_{jm}^{z,\kappa_{l}}
\varphi_{m}^{z,\kappa_{l}},
\end{equation}
where 
\begin{equation}
P^{z,{\kappa_l}}=
\sum_{j}\Big\vert\varphi_{j}^{z,\kappa_{l}}\Big\rangle
\Big\langle\varphi_{j}^{z,\kappa_{l}}\Big\vert
\end{equation}
is the projector onto the space spanned by the
$\varphi_{j}^{z,\kappa_{l}}$ SPFs, $\rho^{z,\kappa_{l}}$ is a density
matrix and $\langle\hat{H}\rangle^{z,\kappa_{l}}$ is a matrix of 
mean-field operators acting on the $\varphi_{j}^{z,\kappa_{l}}$
functions. In its form above, the EOM for the SPFs look identical to 
the usual EOM for the SPFs in the usual MCTDH \cite{bec00:1}. Only the
computation of the density matrices and mean-fields entering the EOM
is now more involved than in a single-layer MCTDH scheme
\cite{wan03:1289,man08:164116,ven11:044135}.

As shown above, the MCTDH and ML-MCTDH EOMs are a set of coupled 
non-linear differential equations, which, however, can be efficiently
solved using standard numerical tools. One may choose {\it propagating}
an initial wave function using a real valued time variable in which
case the physical evolution of the system is modeled. On the basis of 
the same working EOMs, one may use an imaginary time variable in which
case the initial wave function is {\it relaxed} to the ground state of
the Hamiltonian. For MCTDH (but in the Heidelberg package not
for ML-MCTDH) there exist an advanced relaxation version,
which allows to compute excited eigenstates.
Such calculations are called improved relaxation \cite{mey06:179} or 
block improved (BLK) relaxation \cite{dor08:224109}. By a BLK
calculation, a set of initial wave functions is collectively relaxed 
to eigenstates and corresponding energy eigenvalues are obtained at
one time.

\section{Numerical Setup\label{sec:numerical-details}}

\subsection{Hamiltonian Operator}

For representing the potential function and the
wave function, we use primitive grids as detailed in Table
\ref{tab:num-prop}. The definitions of the coordinates (indicated
in the first column) are given in Figure \ref{fig:geom}. We give in 
the second column of Table \ref{tab:num-prop} the primitive basis
functions, which underlay the DVR, together with the number of the
grid points and the range of the grids in atomic units or radian.
Note that the C-O distance is given by $r+r_{\mathrm{min}}$ where 
$r_{\mathrm{min}}=2.1732$ Bohr is the potential minimum along $r$-DOF. 
Similarly, the $z$ coordinate of the CO center-of-mass is
$z+z_{\mathrm{min}}$ where $z_{\mathrm{min}}=3.4771$ Bohr. Thus $r=0$ 
and $z=0$ denote equilibrium positions. In Table \ref{tab:num-prop},
we also give the symbol of the one-dimensional (1D) function for each
coordinate of the initial wave function, as well as the parameters for
these 1D functions, including positions and momenta in the 1D function,
frequency ($\omega_{\mathrm{HO}}$) and mass ($M_{\mathrm{HO}}$) of a 
harmonic oscillator (HO) function, width of a Gaussian function ({\it
i.e.}, variance of the modulus-square of the Gaussian function, 
$W_{\mathrm{GAUSS}}$), and the initial quantum numbers 
$(j_{\mathrm{ini}},m_{\mathrm{ini}})$ of the angular functions.

In Table \ref{tab:mccpd-numerical-details-and-error}, we give the 
numerical details in the MCCPD calculations, including the number 
of trajectory in Monte Carlo calculations and errors of the re-fitting
calculations. The present 6D MCCPD calculation is carried out through 
a total of $\sim8\times10^6$ Monte Carlo points, resulting in 
re-fitting error of $10\;\mathrm{meV}$. On the other hand,
though similar number of Monte Carlo points, the present 21D MCCPD
calculation is carried out, resulting in re-fitting error of about
$48\;\mathrm{meV}$. Noting that the SAP PES was constructed with a 
fitting error of $\sim45\;\mathrm{meV}$ \cite{mar10:074108},
the present re-fitting errors are small enough to obtain reasonable
dynamics results. Finally, the primitive grids used in MCCPD 
are the same as the ones used in the ML-MCTDH calculations,
and are given in Table \ref{tab:num-prop}. In MCCPD we used the
same mode combination scheme as for the wave function,
the scheme is illustrated in Figure \ref{fig:MLtree}.

\subsection{Preparation of initial states\label{sec:initialstates}}

In this work, both the copper surface and the CO molecule are prepared
in the ro-vibrational eigen-state while the translational energy of the
molecule along the $z$ coordinate has, as in all cases, been set to 
$p_z=-17.0$ au, such that the molecule travels towards the surface.
The resulting dynamics calculations are analyzed for initial 
translational energies ranging from $0.0$ to $0.25$ eV. To this
end, total state of the system initially prepared as a product state
\begin{equation}
\vert\Psi(t=0)\rangle=\exp(ip_zz)\,\vert\Psi_{\mathrm{CO}}\rangle\,
\vert\Psi_{\mathrm{surf}}\rangle, 
\label{eq:initialstate}
\end{equation}
where $\vert\Psi_{\mathrm{CO}}\rangle$ and $\vert\Psi_{\mathrm{surf}}\rangle$ 
are vibrational eigenfunctions of CO and Cu(100), respectively, and 
the exponential term accounts for the initial momentum of the CO molecule
towards the Cu surface. To place the CO molecule at some distance from
the surface when preparing $\vert\Psi_{\mathrm{CO}}\rangle$,
we added an artificial harmonic potential with the force constant of
$k=10^{-2}$ au and a minimum at $z=4.0$ Bohr to the $z$-coordinate of
$H_s=T_s+V_s$ and computed the states above as eigenstates of this 
augmented system Hamiltonian, keeping the surface as rigid.
When the artificial harmonic potential is removed, the z-motion of the
CO atiom is no longer in an eigenstate but, after adding the initial
momentum $p_z$, covers an energy range form zero to 0.25 eV. 

One can rationalize the choice of the initial states as follows. At
the distance $z=4.0$ Bohr the interaction with the surface is 
negligible for disturbing the motion of the internal coordinate $r$.
The dipole interaction between CO and surface, however, is already
strong enough to break the rotational symmetry. The CO molecule 
becomes orientated with the C-atom pointing towards the surface. If
one moves the CO molecule adiabatically from infinity to 
$\langle z \rangle=4.0$ Bohr, then GS, AR, and FR correlate with the
free rotational states $(j=0,m=0)$, $(j=1,m=0)$, and $(j=1,m=1)$. The
states GS, BS, AR, and FR were computed using the block-improved 
relaxation algorithm of the Heidelberg MCTDH program. The energies
of the states are reported in Table \ref{tab:results-blk-as-initial-wf}.
From there one can easily find rather good agreement between present 
and previous \cite{men13:164709,men15:164310} results. The differences 
are caused by the different initial $z$ values, $z=4.0$ Bohr in this work
but $z=0.0$ Bohr in our previous calculation \cite{men13:164709,men15:164310}.

As initial states of the CO molecule, $\vert\Psi_{\mathrm{CO}}\rangle$,
we use the ground state (``GS''), the first excited state of the CO 
bond stretch (``BS'') as well as the first excited states associated
with azimuth angle $\phi$ (``AR'') and frustrated zenith angle $\theta$
(``FR'') of a modified system Hamiltonian. 
As initial state of the surface $\Psi_{\mathrm{surf}}(0)\rangle$ we
used the ground state of the bath Hamiltonian $H_b=T_b+V_b$, which is
created by imaginary time propagation of
an initial Hartree product which is close to the ground state. An
approximate excited surface state is subsequently created by multiplying
the $z$-coordinate of the top Cu-atom to the ground state, that is,
\begin{equation} 
\left\vert\Psi_{\mathrm{surf}}^{\mathrm{exc}}\right\rangle=
\frac{Z_1\left\vert\Psi_{\mathrm{surf}}^{\mathrm{GS}}\right\rangle}
{\left\Vert Z_1\left\vert\Psi_{\mathrm{surf}}^{\mathrm{GS}}\right\rangle\right\Vert}.
\label{eq:initialstate-00}
\end{equation}
This creates one quanta of excitation of the out-of-plane mode of the 
central Cu atom. In this work, the out-of-plane excited vibrational 
state of the top atom in the Cu(100) surface is denoted by ``surf''.

\subsection{Analyses\label{sec:flux}}

With the initial state prepared according to equation \eqref{eq:initialstate}
the CO molecule will travel towards the Cu surface where it may either
be reflected and depart again or it is absorbed and sticks to the surface.
The wave function of each initial state is propagate up to $10^4$ fs
by the ML-MCTDH method. To compute the sticking probability of the
molecule on the surface we analyse the flux of the wave function fraction 
through the surface 
positioned at $z=4.0$ Bohr. The sticking probability is obtained from
$1-P_{\mathrm{ref}}(E)$ where $P_{\mathrm{ref}}(E)$ is the probability
of reflection at a given energy $E$. Beyond $z=4.0$ Bohr we place a 
complex absorbing potential (CAP) \cite{ris93:4503,ris96:1409},
\begin{equation}
V_{\mathrm{CAP}}=-i\eta\big(z-z_{\mathrm{CAP}}\big)^n.
\label{eq:cap-potential-0}
\end{equation}
The CAP absorbs the reflected part of the wave packet if it reaches 
the region $z\geq z_{\mathrm{CAP}}$. In the expression of $V_{\mathrm{CAP}}$,
the quantities $n$ and $\eta$ are order and strength of the CAP along
the coordinate $z$, respectively, while $z_{\mathrm{CAP}}$ marks the
starting point of $V_{\mathrm{CAP}}$. In this work, we set $n=3$, 
$\eta=0.005$ au, and $z_{\mathrm{CAP}}=4.5$ Bohr. Furthermore, to 
prevent any reflow of population from the region of the CAP back 
towards the surface, we add a small artificial attractive potential
that sucks a wave packet deeper into the CAP region. For further 
technical details concerning the flux analysis we refer the reader
to reference \cite{men17:184305,jae96:6778,bec00:1}. 

On the other hand, beyond the scattering probability, starting with
the time-evolved wave function $\vert\Psi(t)\rangle$, time-dependent
expectation values 
\begin{equation}
\langle A\rangle=\frac{\langle\Psi(t)\vert A\vert\Psi(t)\rangle}
{\langle\Psi(t)\vert\Psi(t)\rangle}
\label{eq:expectation-values-of-a-00}
\end{equation}
of an observable $A$, are possible. Illustrated in Figures 1 and
2 of the Supporting Information are time-dependent norms
$\langle\Psi(t)\vert\Psi(t)\rangle$ of the present ML-MCTDH wave 
functions. The time-dependent norm becomes less than one for 
for $t>500$ fs, what makes the interpretation of the expectation
values difficult, because the CAP annihilates predominantly the
fast-moving long-ranged parts of the wave packet. We therefore
show expectation values only for times up to $500$ fs.
In this work, we used the following observable operators:
the CO total energy $H_s=T_s+V_{SAP}(\mathbf{q},0)$,
the surface energy, $H_b=T_b+V_b$, the distance of CO from the
surface, $Z$, and the out-of-plane motion of the top Cu-atom, $Z_1$.

\section{Results and Discussions\label{sec:results}}

\subsection{Rigid Surface\label{sec:rigid-surface-flux}}

\begin{figure}[h!]
  \centering
   \includegraphics[width=16cm]{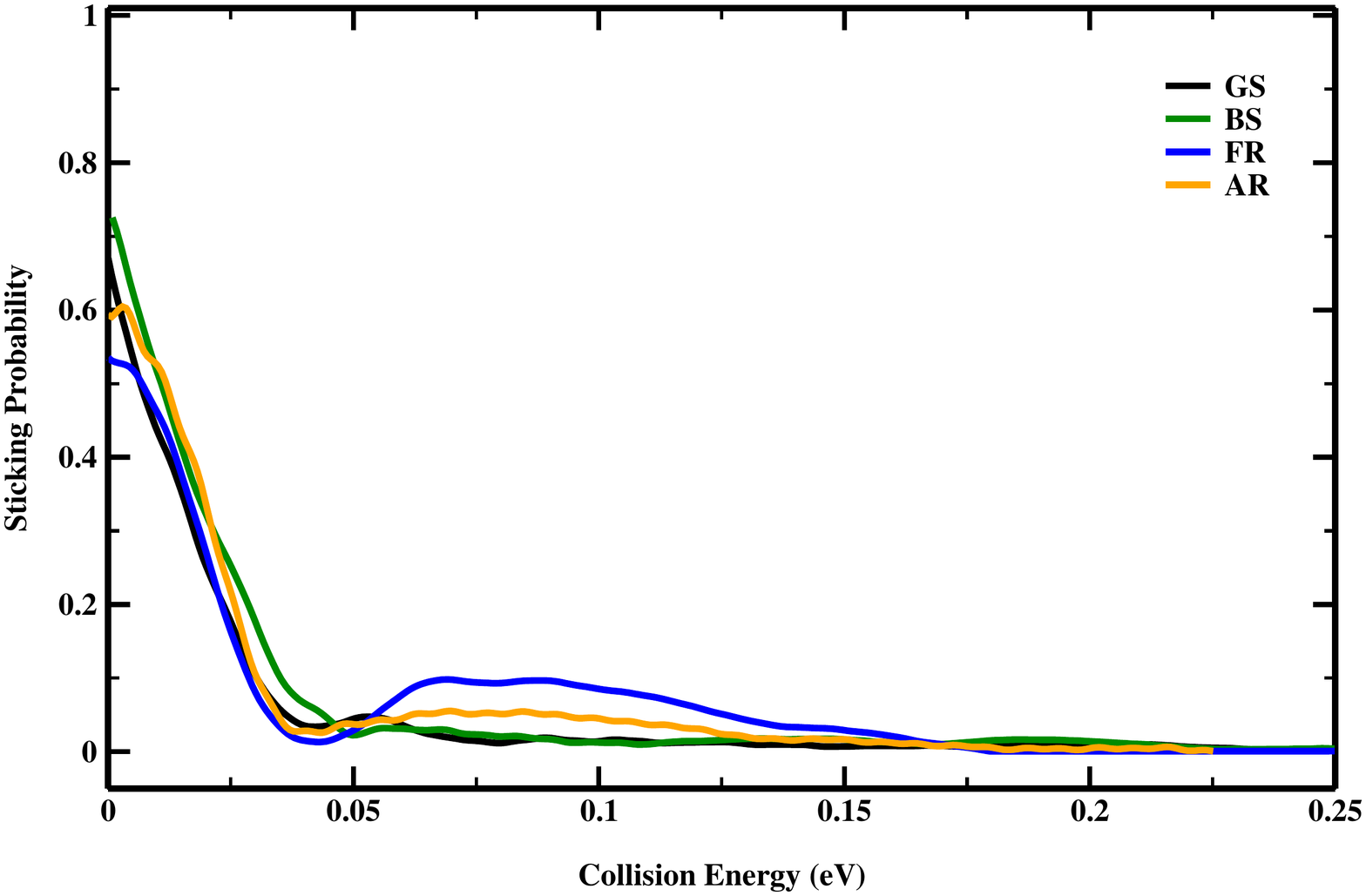}
    \caption{\figfoot}
     \label{fig:flux-6d0d}
      \end{figure}

With the various initial initial states as discussed in Section
\ref{sec:initialstates}, extensive ML-MCTDH propagations and 
follow-up flux and expectation analyses (see Section \ref{sec:flux})
are performed to compute the sticking probabilities. In this section,
we shall show results obtained with the 6D Hamiltonian model, 
{\it i.e.} scattering off a rigid surface. Figure \ref{fig:flux-6d0d} 
displays the sticking probabilities as a function of the collision
energies for the different initial states. One observes that the 
sticking probabilities drop quickly to small values for the collision 
energies above $0.03\;\mathrm{eV}$ ($\sim240\;\mathrm{cm}^{-1}$). For
the rotational excited states, AR and FR, there is a small recovery 
of the sticking between $0.03$ and $0.15\;\mathrm{eV}$, but above 
$0.15\;\mathrm{eV}$ almost all incoming particles are reflected. 
Since in the present model the Cu(100) surface is rigid, the collision
energy cannot be transferred to surface DOFs. To accomplish sticking,
the incoming translational energy has to be partially transferred to 
$xy$-motion (T mode), or to $\theta\phi$-motion (frustrated rotation,
R mode). 

\begin{figure}[h!]
  \centering
   \includegraphics[width=16cm]{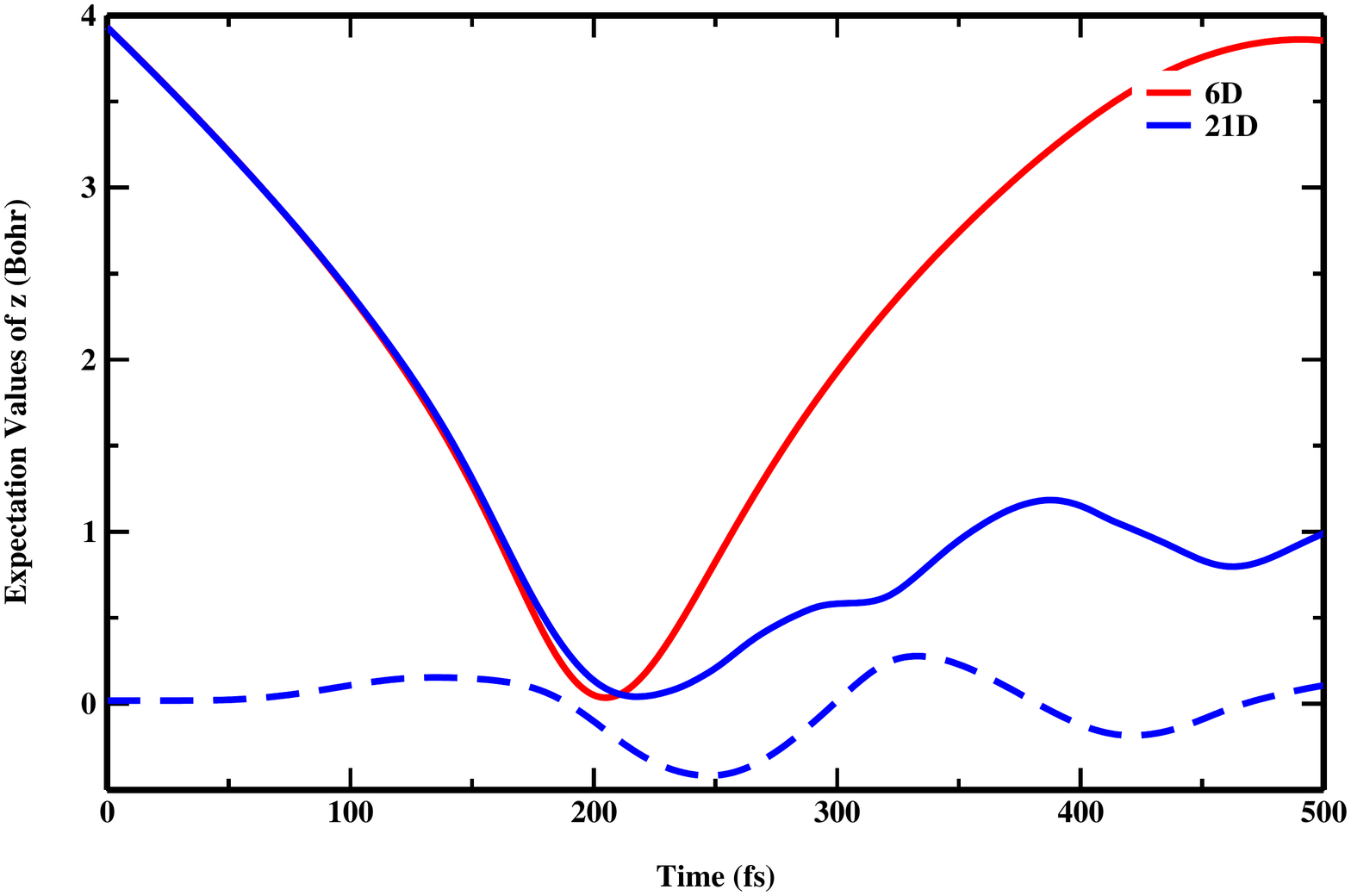}
    \caption{\figfoot}
     \label{fig:expectation-compar}
      \end{figure}

\begin{figure}[h!]
  \centering
   \includegraphics[width=16cm]{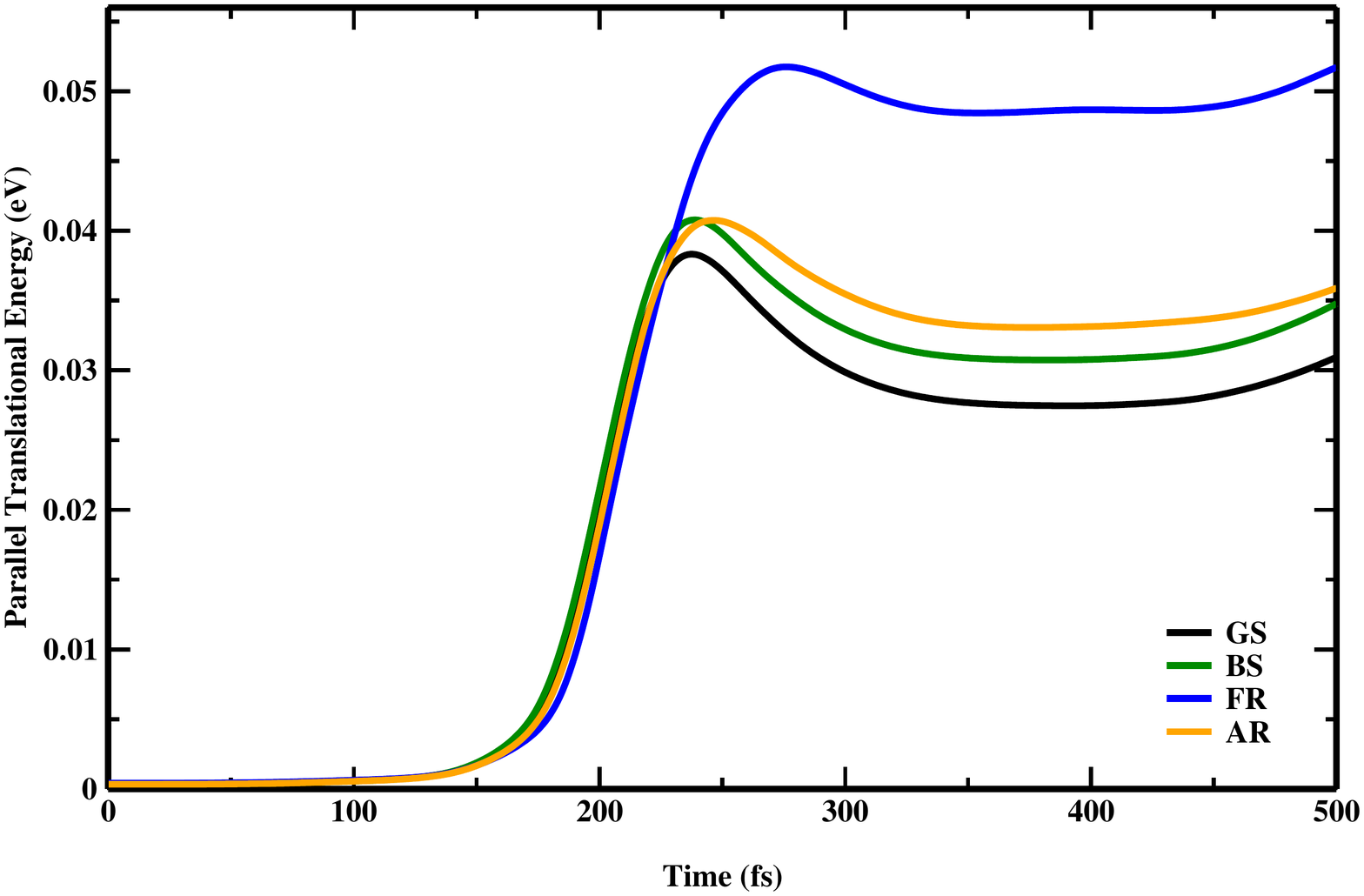}
    \caption{\figfoot}
     \label{fig:xy-6d0d}
      \end{figure}

When the CO molecule approaches the surface, it is accelerated by
about $0.4\;\mathrm{eV}$ due to the attractive part of the interaction 
potential. Near and slightly above $t=200$ fs the wavepacket is closest
to the surface (compare with Figure \ref{fig:expectation-compar}). Figure 
\ref{fig:xy-6d0d} shows the kinetic energy of the $xy$-motion for four
initial states versus time. Near $t=200$ fs the energy rises
sharply, then there is a small back-transfer of energy, and in the end
between $27$ and $48$ meV are transferred to motion parallel 
to the surface. The increase of the expectation values for $t>450$ fs
is due to the decrease in norm as discussed above.

The energy transfer is smallest, if the CO molecule is
initially in its ro-vibrational ground state. The elongation of the C-O 
bond, due to a vibrational excitation, helps to enhance the energy 
transfer to parallel motion. But more efficiently this energy transfer
is enhanced by initial rotational excitations. The CO rotational energy
over time is shown in Figure 5 of the Supporting Information. During 
the collision there is a virtual rotational excitation of 
$15\sim30\;\mathrm{meV}$, but the final energy transfer is smaller than
$10\;\mathrm{meV}$. Again, initial rotational excitation enhances the 
energy transfer, but it remains small. Hence the sticking on a rigid 
surface is mainly due to energy transfer to parallel motion. 

\subsection{Flexible Surface\label{sec:movable-surface-flux}}

\begin{figure}[h!]
  \centering
   \includegraphics[width=16cm]{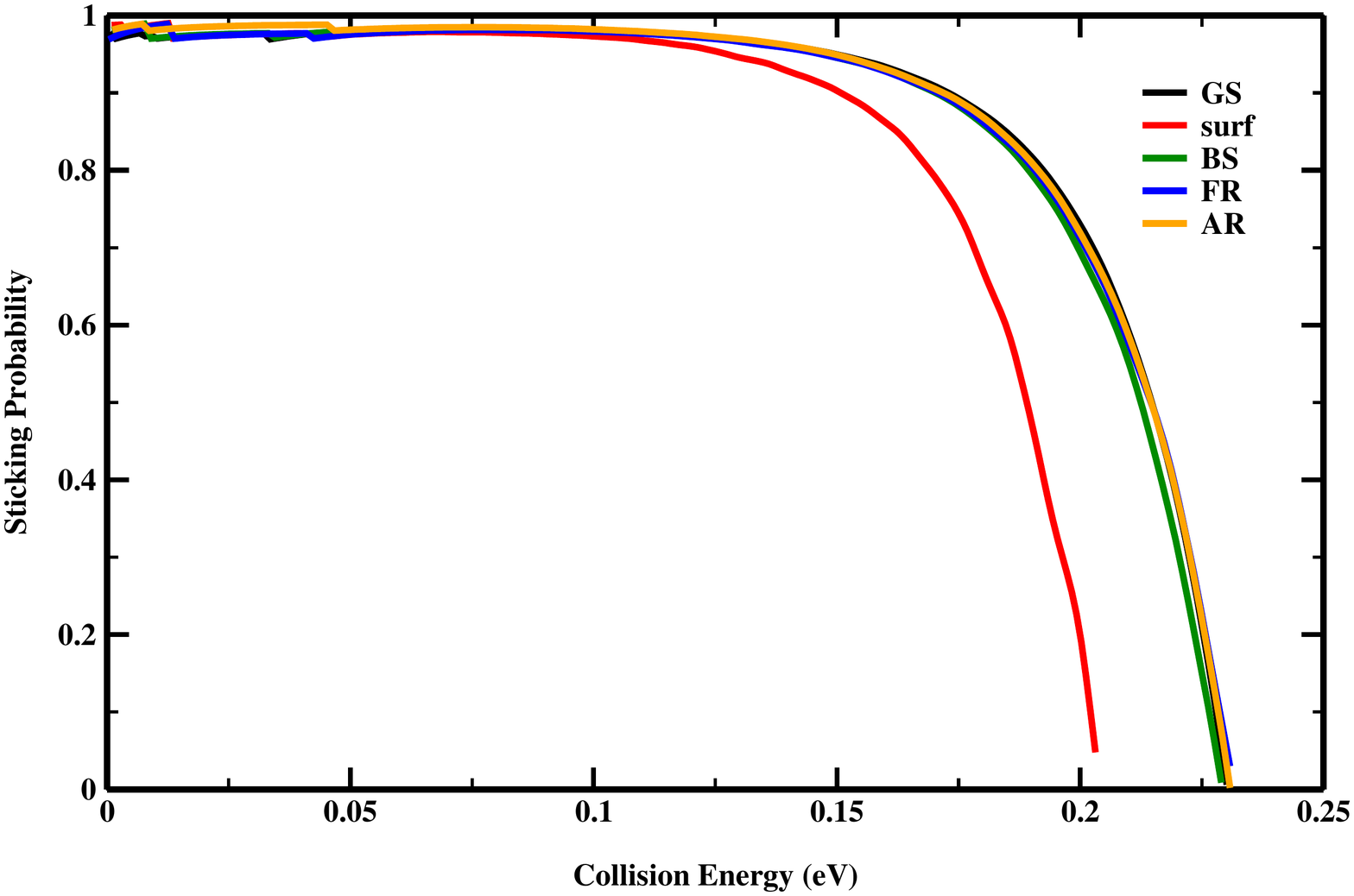}
    \caption{\figfoot}
     \label{fig:flux-21d}
      \end{figure}

In Figure \ref{fig:flux-21d} is shown the sticking probabilities
of the 21D results from the direct model. The sticking probabilities
decrease much more slowly compared to the case of rigid surface 
discussed in Section \ref{sec:rigid-surface-flux}, but they start
to decay quickly for collision energies above $0.15$ eV.

The figure shows sticking probabilities, which separate into two 
categories. First, an excitation of the out-of-plane mode of the
center atom attenuates the sticking process, and the corresponding
21D sticking probability (red line, denoted by  ``surf''). has the 
smallest values. Obviously, a vibrational excitation of surface 
atoms helps to kick-off the CO molecule from the surface. Second,
the sticking probabilities computed for vibrationally or rotationally
excited  states of the CO molecule  are close to each other.
The initial ro-vibrational state of CO is less relevant.

\begin{figure}[h!]
  \centering
   \includegraphics[width=16cm]{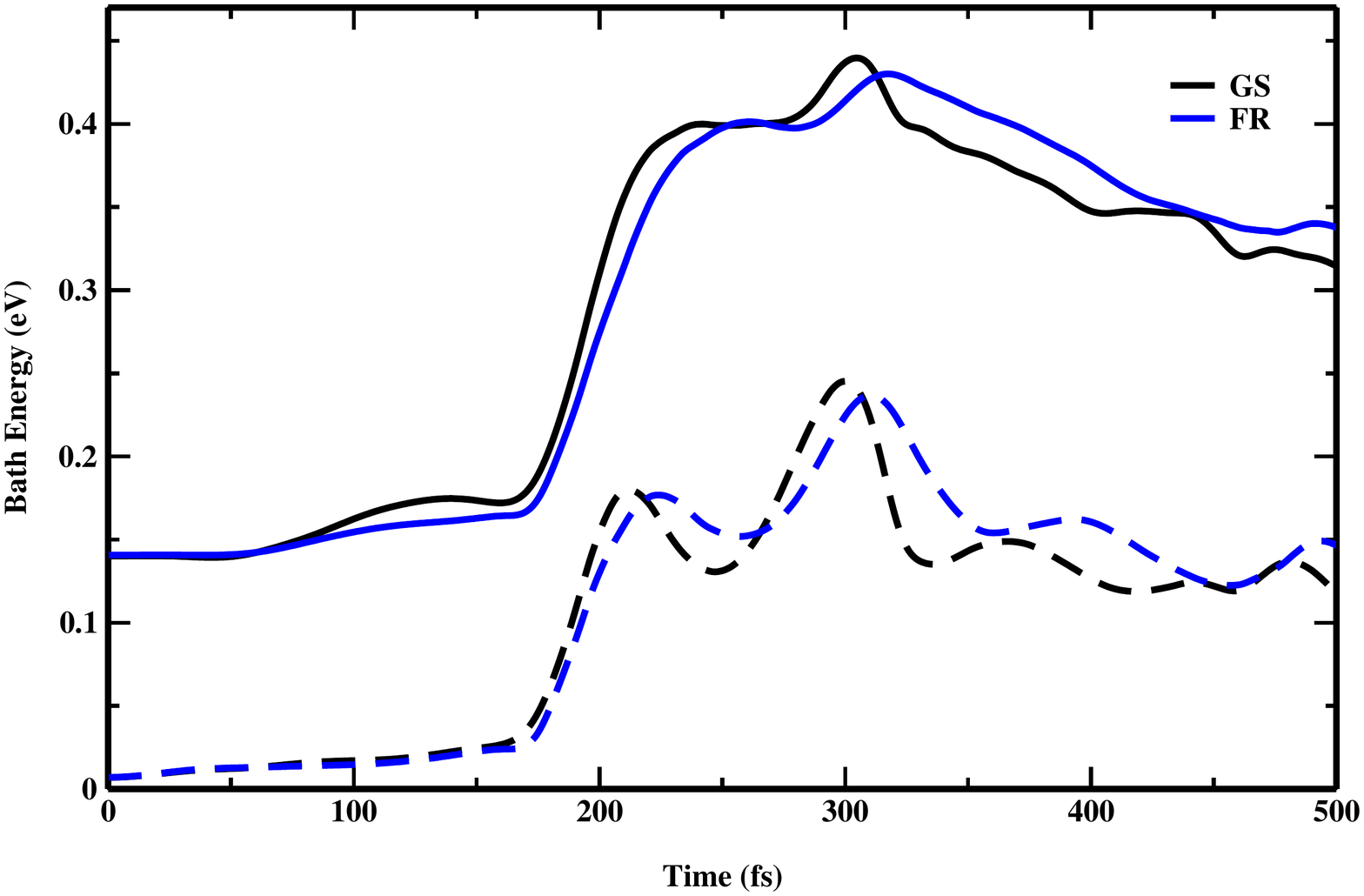}
    \caption{\figfoot}
     \label{fig:bath}
      \end{figure}

The importance of the bath is demonstrated by Figure \ref{fig:bath}, 
which shows the expectation energy of the bath Hamiltonian versus 
collision time. During the collision the bath absorbs about 
$0.21$ eV, which is much more compared to the energy going 
into parallel motion (compare with Figure \ref{fig:xy-6d0d}). This 
large energy transfer explains the large sticking probability. The
two lower lines in Figure \ref{fig:bath} display the energy expectation
value of the out-of-plane motion of the top Cu-atom. This motion absorbs 
about $0.13$ eV, that is more than half of the energy 
transferred to the bath is taken by the out-of-plane motion of the
top Cu-atom.

\subsection{Discussions\label{sec:discuss-flux}}

\begin{figure}[h!]
  \centering
   \includegraphics[width=16cm]{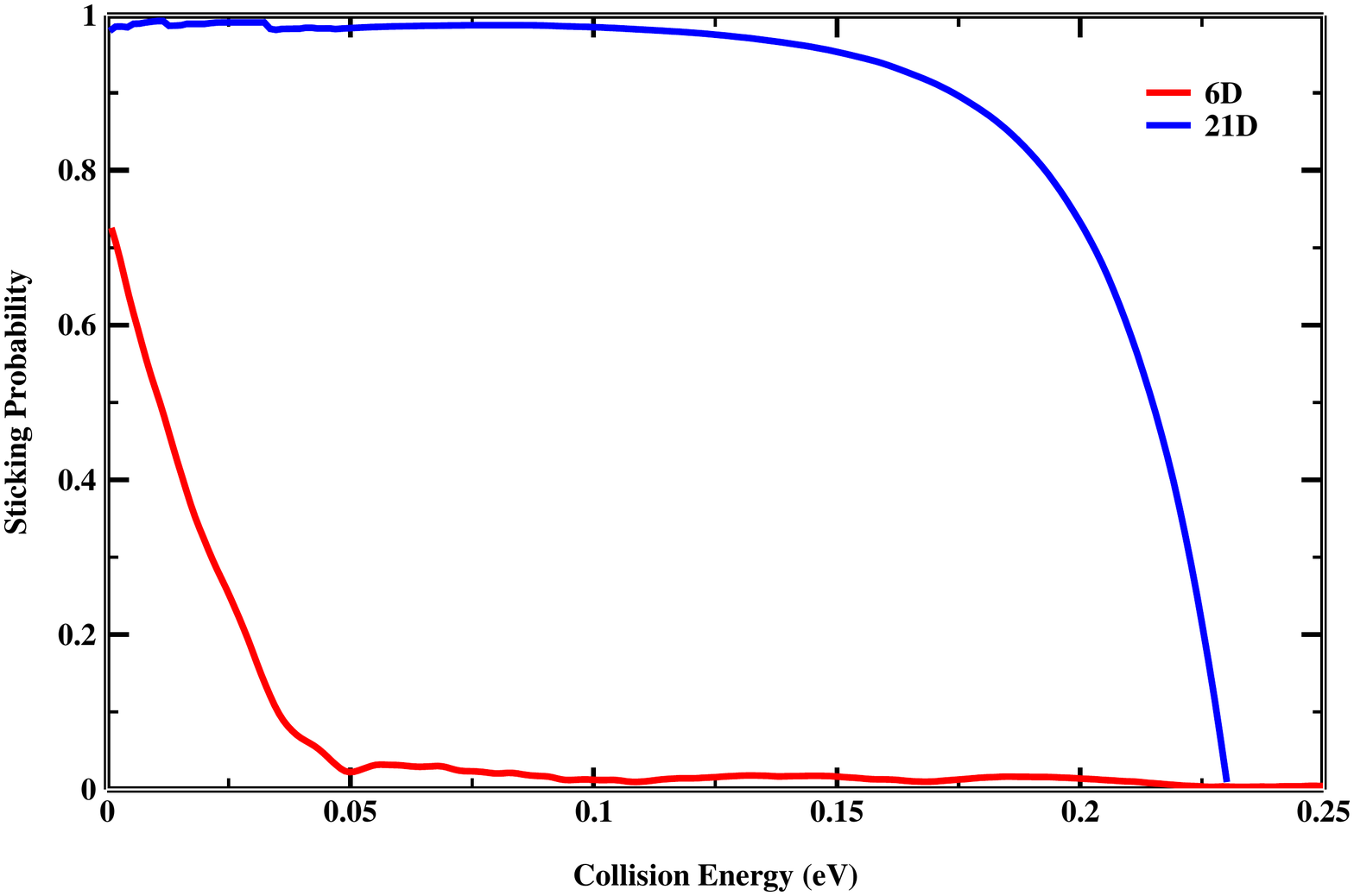}
    \caption{\figfoot}
     \label{fig:flux-compar}
      \end{figure}

In Figure \ref{fig:flux-compar} the computed sticking probabilities
of the CO molecule in its ro-vibrational ground state are compared for
the model with rigid and with flexible surface atoms, respectively.
As shown in Figure \ref{fig:flux-compar}, at collision energies
below $0.15$ eV the sticking probability of the model with 
flexible surface atoms shows a very slow decrease, in contrast to
the case where all surface atoms are fixed. Further increasing the
collision energy to above $0.15$ eV let the sticking probability 
for flexible surface atoms start to decrease and to reach zero when 
the collision energy is above $0.20$ eV. In any case, the 
sticking probability is much larger when the surface atoms are treated
as flexible atoms as compared to the case when all surface atoms are
fixed. This difference is caused by the surface DOFs that let the CO 
molecule transfer translational kinetic energy to other modes, in
particular to the bath modes. Hence the surface DOFs play an important
role in surface scattering. 

To visualize the dynamics behavior of the surface scattering of CO,
we show the time-dependent expectation values of the $z$-coordinate
of CO and (if present) of the top Cu atom in Figure \ref{fig:expectation-compar}.
In the case of a rigid surface (the black line), the CO molecule
collides with the surface near $200$ fs, where the position
of CO almost reaches $z=0$. Remember that $z=0$ is the equilibrium 
position of a CO atom adsorbed on the surface. As the displayed value
is an expectation value averaged over the wavepacket, large parts of 
the wavepacket have entered the region $z<0$ where a strong repulsion
sets in. The wavepacket is thus reflected and the $z$ value increases
again. Above $t=400$ fs the $z$ curve flattens and starts to turn back.
This is an artifact caused by the CAP absorption, which sets in at 
$t=400$ fs. As the distant parts of the wavepacket are annihilated
by the CAP, the average $z$ value is reduced.

In the case of a flexible surface (blue lines), the situation changes.
Since the attractive interaction between the center and flanking copper
atoms, the center atom slightly moves upwards and is then pushed downwards
by the hard collision with CO. The $z$ turning point occurs at about
$220$ fs, which is slightly larger compared to the collision
with a rigid surface. After the collision the $z$ expectation value 
increases much slower as in the case of a rigid surface. This is because
a large fraction of the wavepacket remains close to the surface. Above
$400$ fs the expectation value is again distorted by the
absorption due to the CAP. After the collision, the center surface atom
performs an oscillatory motion, which is slightly damped due to the 
coupling to the other surface atoms. Near $t=330$ fs, when 
the $Z$ coordinate assumes its maximum value, the center atom pushes
out some of the part of the wavepacket, which has remained close to 
the surface. This back-transfer of bath energy, which can also be 
observed in Figure \ref{fig:bath}, reduces the sticking probability.

\section{Conclusions\label{sec:con}}

To study mode-specific features in surface scattering of CO/Cu(100) 
with lattice effects, a surface model with five flexible copper atoms
is employed and the ML-MCTDH method is used to perform the quantum
dynamics calculations. To perform the ML-MCTDH calculations efficiently,
the potential terms are firstly re-fitted to the CPD form using the 
Monte Carlo method by the MCCPD method \cite{sch20:024108}. 

The first few ro-vibrational states of CO fixed at a distance from the
surface are computed and serve as initial states of the  ML-MCTDH 
propagations. On the basis of these calculations, we find that the
sticking probabilities are strongly affected by the inclusion of 
movable surface atoms, but depend only weakly on initial ro-vibrational 
states of CO. The sticking probability is reduced, it the out-of-plane
motion of the top atom is initially excited, compared to a surface in
its vibrational ground state. With the aid of the time-dependent 
expectation values the scattering process is further analyzed. 

\section*{Supplementary Material}
 
See Supplementary Material Documents at http://dx.doi.org/XXX for
details of the results of the expectation analyses.

\section*{Acknowledgments}

Q.M. gratefully acknowledges financial support by National Natural Science
Foundation of China (Grant No. 21773186), National Natural Science Foundation
of Shaanxi Province (Grant No. 2019JM-380), Fundamental Research Funds for the
Central Universities (Grant No. 3102017JC01001), and Hundred-Talent Program of 
Shaanxi. M.S. and H.-D.M. gratefully acknowledges financial support by the {\it
Deutsche Forschungsgemeinschaft} (DFG) project ME623/22-1.
The authors wish to thank Prof. Dr. R. Marquardt (Strasbourg, France)
for providing us their FORTRAN routine for the SAP PES.



\clearpage
\begin{sidewaystable}
 \caption{%
DVR-grids used in the dynamics calculations, and parameters of the 
initial wave function for the ML-MCTDH propagation calculations. The
definitions of the coordinates (indicated in the first column) are 
given in Figure \ref{fig:geom}(b) as well as Figure 1 of reference 
\cite{men15:164310}. The second column describes the primitive basis
functions, which underlay the DVR. The third column gives the number
of the grid points. The fourth column gives the range of the grids 
in atomic unit or radian. Note that the C-O distance is given by
$r+r_{\mathrm{min}}$ where $r_{\mathrm{min}}=2.1732$ Bohr is the
potential minimum along $r$-DOF. Similarly, the $z$ coordinate 
of the CO center of mass is $z+z_{\mathrm{min}}$ where
$z_{\mathrm{min}}=3.4771$ Bohr. Thus $r=0$ and $z=0$ denote 
equilibrium positions. The fifth column gives the symbol of
the one-dimensional (1D) function for each coordinate of the
initial wave function. The other columns give the parameters
for these 1D functions, including positions and momenta in the
1D function, frequency ($\omega_{\mathrm{HO}}$) and mass
($M_{\mathrm{HO}}$) of the HO function, width of the GAUSS 
function ({\it i.e.}, variance of the modulus-square of the 
Gauss function, $W_{\mathrm{GAUSS}}$), and the initial quantum
numbers $(j_{\mathrm{ini}},m_{\mathrm{ini}})$ of the angular 
functions.
}%
\begin{center}
 \begin{tabular}{llllllrlrlrlrlr}
  \hline
 Coordinates $^a$ &~~& \multicolumn{5}{c}{Primitive basis function}  &~~~~~~& \multicolumn{7}{c}{Initial wave function}  \\ \cline{3-7} \cline{9-15}
                  &~~& Symbol $^b$ &~~& Grid points &~~& Range of the grids &~~~~~~& Symbol $^c$ &~~& Position &~~& Momentum &~~& Parameters  \\
\hline
\multicolumn{15}{l}{{\it Coordinates of the CO molecule}, $\mathbf{q}$}      \\
$x$ and $y$  &&  EXP   && 45 && $[-2.415,2.415]$ && HO    && $0.0$  && $0.0$  &&
$\omega_{\mathrm{HO}}=31.8$ cm$^{-1}$, $M_{\mathrm{HO}}=28.0$ AMU   \\
$z$          &&  FFT   && 192 && $[-1.000,6.000]$ && GAUSS && $4.0$  && $-17.0$
&& $\mathrm{width}=0.1$   \\
$r$          &&  HO    && 21  && $[-0.300,0.300]$ && EIGENF&& --  && --  && ground state \\
$\theta$     &&  PLEG  && 45 && $[0,\pi]$        && KLEG  && ---    && ---    && $j_{\mathrm{ini}}=j$ \\
$\phi$       &&  EXP   && 27 && $[0,2\pi]$       && K     && ---    && ---    && $m_{\mathrm{ini}}=m$ \\
\hline
\multicolumn{15}{l}{{\it Coordinates of the Cu atoms}, $\mathbf{Q}_i$}  \\
$Q_{1z}$      &&  SIN  && 55 && $[-0.800,1.200]$ && HO && $0.0$  && $0.0$  && $\omega_{\mathrm{HO}}=103.4$ cm$^{-1}$, $M_{\mathrm{HO}}=63.546$ AMU \\
$Q_{i\alpha}$ &&  HO  && 15  && $[-0.397,0.397]$ $^d$ && HO && $0.0$  && $0.0$  && $\omega_{\mathrm{HO}}=103.4$ cm$^{-1}$, $M_{\mathrm{HO}}=63.546$ AMU \\
$Q_{jz}$      &&  HO  && 15  && $[-0.397,0.397]$ $^d$ && HO && $0.0$  && $0.0$  && $\omega_{\mathrm{HO}}=103.4$ cm$^{-1}$, $M_{\mathrm{HO}}=63.546$ AMU \\
\hline
 \end{tabular}
  \end{center}
   \label{tab:num-prop}
    \end{sidewaystable}
\clearpage
\quad  \\
$^a$ Indices in $Q_{i\alpha}$ are $i=1,3,5,6,8$ and $\alpha=x,y$, 
while index in $Q_{jz}$ is $j=3,5,6,8$.   \\
$^b$ ``EXP'', ``SIN'', and ``HO'' stand for exponential, sine, and 
harmonic oscillator DVR, respectively. ``FFT'' denotes fast Fourier
transform collocation. ``PLEG'' denotes a two-dimensional extended
Legendre DVR \cite{suk01:2604} for angular coordinates. \\
$^c$ ``HO'' and ``GAUSS'' designate the choice of harmonic oscillator
eigenfunction and Gaussian function, respectively, as initial SPFs.
EIGENF means eigenfunction of a specified potential which, in this 
work, is SAP potential along $r$ setting $z$ is sufficiently large 
($z=4.0$ Bohr) as to make the interaction between CO and Cu(100)
negligible. ``KLEG'' and ``K'' denotes associated Legendre function
and body-fixed magnetic quantum number, respectively, to specify the
initial wavefunction. \\ 
$^d$ In these cases only the position (that is $0.0$) and frequency
(that is $103.4\;\mathrm{cm}^{-1}$) parameters of the HO DVR are 
given, and then the ranges of the grids are calculated automatically.

\clearpage
 \begin{table}
  \caption{%
Numerical details of the present MCCPD calculations \cite{sch20:024108}.
The first column gives the method used in the present Monte-Carlo
sampling. The second column gives the temperature ($k_{\mathrm{B}}T$ in eV) of each kind
Metropols sampling. These samplings are distributed according to the weight
$w(\mathbf{q},\mathbf{Q})=\exp(-V_{\mathrm{SAP}}(\mathbf{q},\mathbf{Q})/(k_{\mathrm{B}}T))$.
The third and fourth columns present the numbers
of the sampling points used for the fit and test, respectively.
For the CPD fit the combined set of  sampling points of the four distributions given in 
the \texttt{FIT} column are used.
The computation of the fit error is performed for each temperature
distribution individually, and for all test points all together. 
Note that fit- and test-distributions are independent.
The fifth and sixth columns give energy
errors (in meV) between refitted and original potential using 
text sampling points, including their average (denoted by \texttt{MEAN})
and root-mean-square (denoted by \texttt{RMS}) values, respectively. 
}%
\begin{tabular}{lclccccccccccc}
 \hline
     &~~& \multicolumn{7}{c}{Trajectory} &~~~~&\multicolumn{3}{c}
     {Testing Error (meV)} \\
\cline{3-9}\cline{11-13} 
&& Method &~~& $k_{\mathrm{B}}T$ (eV) &~~& \texttt{FIT} &~~& \texttt{TEST}
 &~~& \texttt{MEAN} &~~& \texttt{RMS} \\

\hline
\multicolumn{13}{l}{{\it 6D PES}}  \\
         && Metropolis  && $0.05$ && $6.00\times10^5$ && $1.00\times10^6$ && $-3.58\times10^{-4}$ && $0.50$ \\
         && Metropolis  && $0.25$ && $3.00\times10^5$ && $1.00\times10^6$ && $-2.22\times10^{-2}$ && $1.75$\\
         && Metropolis  && $0.50$ && $1.00\times10^5$ && $1.00\times10^6$ && $-3.21\times10^{-3}$ && $4.12$ \\
         && Metropolis  && $1.00$ && && $1.00\times10^6$ && $3.14\times10^{-2}$ && $9.77$ \\
         && Metropolis  && $1.50$ && && $1.00\times10^6$ && $2.52\times10^{-2}$ && $12.44$ \\
         && Metropolis  && $2.50$ && && $1.00\times10^6$ && $4.02\times10^{-3}$ && $14.11$ \\
         && uniform  &&        && $5.00\times10^4$ && $1.00\times10^6$ && $-2.26\times10^{-2}$ && $17.79$  \\
{\it in total}&& && &&$1.05\times10^6$&&$7.00\times10^6$&&$1.76\times10^{-3}$&&$10.06$\\
\hline
\multicolumn{13}{l}{{\it 21D PES}}  \\
         && Metropolis  && $0.05$ && $2.00\times10^5$ && $1.00\times10^6$ && $0.36$ && $18.30$  \\
         && Metropolis  && $0.25$ && $1.05\times10^5$ && $1.00\times10^6$ && $0.15$ && $29.08$ \\
         && Metropolis  && $0.50$ && $5.00\times10^4$ && $1.00\times10^6$ && $1.03$ && $33.44$  \\
         && Metropolis  && $1.00$ &&  && $1.00\times10^6$ && $0.58$ && $47.61$   \\
         && Metropolis  && $1.50$ &&  && $1.00\times10^6$ && $0.15$ && $54.96$   \\
         && Metropolis  && $2.50$ &&  && $1.00\times10^6$ && $-0.37$ && $60.95$ \\
         && uniform     &&   && $5.00\times10^4$ && $1.00\times10^6$ && $-2.02$ && $70.92$ \\
{\it in total}&& && &&$4.50\times10^5$&&$7.00\times10^6$&& $-1.63\times10^{-2}$ &&$48.30$ \\
\hline
 \end{tabular}
  \label{tab:mccpd-numerical-details-and-error}
   \end{table}

\clearpage
 \begin{table}
  \caption{
Excitation energies of the first excited states (all energies in
cm$^{-1}$) of the CO bond stretch (denoted by BS) as well as the
Azimuth and frustrated rotation (denoted by AR and FR) based on 
the MCTDH relaxation calculations, where the initial CO molecule 
is located at $z=4.0$ au. The excitation energies of the overtone
states of these modes are also given. For comparison, the previous
excitation energies for the CO molecule adsorbed on the surface are
also shown. The first and second columns give the vibrational modes
and corresponding symbols of the system. The third column present 
the present MCTDH relaxation results at the 6D $\{x,y,z,r,\theta,\phi\}$
level. The other three columns give the previous theoretical and 
experimental results for the excitation energies of CO adsorbed
on the surface.
}
\begin{tabular}{lllccccccccc}
 \hline
&~~&&\multicolumn{3}{c}{$\langle z\rangle=4$ Bohr $^a$}&~~&\multicolumn{5}{c} {$\langle z\rangle=0$, adsorbed$^b$}  \\
     \cline{4-5}\cline{8-12}
Mode &~~& Symbol &~~ & MCTDH  &~&~&  MCTDH $^c$ &~~& Lanczos $^d$ &~~& Expt. $^e$ \\
\hline
C-O stretch      && BS && $2141.2$ &~&& $2046.5$ && $2052.5$ && $2079.3$ \\
Azimuth rotation && AR && $48.2$   &~&&   ---    && ---      && ---  \\
Frustrated rotation&&FR&& $92.2$   &~&& $366.1$  && $366.1$  && $345.2$ \\
C-O stretch with azimuth rotation&&BS + AR&& $2189.4$ &~&& --- && --- && ---  \\
C-O stretch with Frustrated rotation&&BS + FR&&$2233.2$&~&&$2412.6$ &&$2418.6$ &&$2424.5$ \\
\hline
 \end{tabular}
  \label{tab:results-blk-as-initial-wf}
   \end{table}
\quad \\
$^a$ The present work, where the CO molecule is located at $\langle 
z\rangle=4.0$ au with the aid of an augmented harmonic potential
(see the maintext).  \\
$^b$ The previous work, where the CO molecule is adsorbed on the 
Cu(100) surface.  \\
$^c$ Reference \cite{men13:164709}.   \\
$^d$ Reference \cite{mar10:074108}.  \\
$^e$ Reference \cite{hir90:480,ell95:5059,gra98:7825}.

\clearpage
\section*{Figure Captions}

\figcaption{fig:geom}{%
(a) Definition of the atomic arrangement of the CO/Cu(100) system. The 
red and gray circles represent the oxygen and carbon atoms, respectively,
and the pink circles represent the copper atoms. Coordinates of CO, center
copper atom (called Cu1), and four non-center copper atoms (called Cu3, 
Cu5, Cu6, and Cu8) nearest to Cu1 are involved in calculations. These 
five copper atoms are selected from a larger nine-by-nine atoms grid,
in which the copper atoms are ordered in three planes (layers) 
perpendicular to the $z$-axis. For clarity, the four second-nearest
copper atoms (called Cu2, Cu4, Cu7, and Cu9) in the surface layer are
also shown. In subfigure (b) we show the definitions of the 
system-coordinates in this work. The origin of the Cartesian coordinate 
system is the equilibrium position of the top atom Cu1. Here $\{x,y,z\}$
are the Cartesian coordinates of the center-of-mass (denoted by G) of 
the CO molecule. Variable $r$ is the bond length of the CO molecule. 
Variable $\theta$ is the polar angle between C-O bond and the $z$ axis,
whereas $\phi$ is the azimuthal angle. The figure is reprinted with 
permission from {\it J. Chem. Phys.} {\bf 146}, 184305 (2017). Copyright
2017 American Institute of Physics.
}%

\figcaption{fig:MLtree}{%
The ML-tree used in  the 21D calculations.
Each circle represents a node and the squares represent the primitive basis
functions, i.e. the grids. The numbers on the lines between the circles
represent the numbers of SPFs used on the node, and the numbers
between a circle and a square are the number of grid points used for
the particular DOF. This is a four layer tree ($l$=4).
}%

\figcaption{fig:flux-6d0d}{%
Sticking  probabilities versus collision energy (in eV) computed for
the  6D model which has coordinates $\{x,y,z,r,\theta,\phi\}$.
The black and red lines represent the sticking probabilities when
CO is initially in its ground state (denoted by GS) and in the first
vibrationally excited state of the C-O stretch mode (denoted by BS).
The blue and green lines represent those of excited states
associated with frustrated azimuth angle $\phi$ (denoted by FR)
and zenith angle $\theta$ (denoted by AR), respectively.
}%

\figcaption{fig:expectation-compar}{%
Comparison of the time-dependent position expectation values along the $z$
coordinate of CO (the solid lines) and the top copper atom (the dashed
line). The red and blue lines represent the results from 6D (i. e. rigid surface)
and 21D models, respectively. The initial ro-vibrational state of
CO is its ground state.
}%

\figcaption{fig:xy-6d0d}{%
Kinetic energy of the motion parallel to the surface versus collision time.
The four lines refer to different ro-vibrational initial states of CO.
A rigid surface is used (6D model). The increase of the energy for
times $t> 400$ fs is an artifact. The wavepacket reaches the CAP at
$t=400$ fs and the norm decreases sharply thereafter. The parts of the
wavepacket which are close to the surface get a higher weight when
evaluating the expectation value for times larger than 400 fs.
In general, it is difficult to interpret expectation values after CAP
absorption has set in.
}%

\figcaption{fig:flux-21d}{%
Sticking  probabilities versus collision energy (in eV) computed for
the  21D model with flexible surface atoms.
The black and green lines represent the sticking probabilities
when the CO  molecule is initially in the ground state (denoted by GS)
and vibrational excited state of the CO bond stretch mode
(denoted by BS), the yellow and blue lines represent those where the
molecule is in the excited states associated with frustrated azimuth
angle $\phi$ (denoted by AR) and zenith angle $\theta$ (denoted by FR),
respectively. The red line represents the sticking probability 
when the molecule is initially in the ground state but the surface is
in the excited state associated with the out-of-plane mode of the
center atom (denoted by surf).
}%

\figcaption{fig:bath}{%
Energy of the bath versus collision time. Initially the CO molecule
is in its ro-vibrational ground (black line) state or in the rotationally
excited FR state (blue line). The two dashed lines show the energy
expectation value of the Z-motion of the top Cu-atom.
}%

\figcaption{fig:flux-compar}{%
Comparison of the sticking probabilities for the 6D (red line),
21D (blue line), and 21D (green line) models, where all sticking
probabilities are computed with CO initially in its ground state.
}%

\clearpage

\clearpage

\clearpage

\clearpage

\clearpage

\clearpage

\clearpage

\clearpage

\end{document}